\documentclass[a4paper,12pt]{article}
\usepackage{latexsym}
\usepackage{exscale}
\usepackage{graphicx}
\usepackage[USenglish]{babel}
\usepackage{amsmath}
\usepackage{amssymb}
\begin{document}
%
%
\title{\bf\sf Magneto-optical properties of $\mathrm{Co} \! \mid
\!\mathrm{Pt}$ multilayer systems}
\author{
{\sf A.~Vernes}$^{\, a)}$,
{\sf L.~Szunyogh}$^{\, a,b)}$, {\sf and P.~Weinberger}$^{\, a)}$ \\
$^{a)}$ {\small \it Center for Computational Materials Science}, 
 {\small \it Technical University Vienna}, \\
 {\small \it Gumpendorferstr. 1a, 1060 Vienna, Austria} \\
$^{b)}$  {\small \it Department of Theoretical Physics }, 
 {\small \it Budapest University of Technology and Economics}, \\
 {\small \it Budafoki \'{u}t 8, 1521 Budapest, Hungary}}
\renewcommand{\today}{}
\maketitle
%
%
%
%
\begin{abstract}
We are reporting, for the first time in the literature, theoretical
Kerr spectra of $\mathrm{Co} \!  \mid\!\mathrm{Pt}$ multilayer systems
as obtained on a first principles basis including multiple reflections
and interferences from all the boundaries in--between the layers.
\end{abstract}
%
%
\section{Introduction}
\label{sect:intro}
%
$\mathrm{Co} \! \mid\!\mathrm{Pt}$ multilayer systems are thought to
be the next--generation magneto--optical recording media, because
their performance is very similar to that of the rare--earth based
materials, which are already in use$^{\,1}$. Although, in the last
decade a large amount of experimental investigations has been
performed on these systems, realistic theoretical investigations, are
still lacking up to now.

The propagation of electromagnetic waves in any multilayer system can
exactly be described by using either a $2\times2$ matrix$^{\,2}$ or
$4\times 4$ matrix$^{\,3}$ formalism.  However, according to our
knowledge, up to date, these techniques have been only used to
simulate magneto--optical properties of multilayers using exclusively
bulk optical data from experiments. The present contribution completes
these kind of theoretical investigations. Here the complex Kerr effect
is calculated on a first principle basis by using the theoretical
layer--resolved optical conductivity tensor of the investigated
layered system as input for the $2\times2$ matrix formalism appropiate
for polar geometry at normal incidence.
%
\section{Theoretical framework}
\label{sect:theory}
%
The complex optical conductivity tensor is calculated using
Luttinger's formula$^{\,4}$ by means of a contour integration
technique$^{\,5}$, which permits the computation to be performed at
nonzero temperatures and for finite life--time broadening.  In
combination with the spin--polarized relativistic screened
Kor\-ringa--Kohn--Rostoker band structure method$^{\,6}$, this
technique has been shown to provide the most adequate first principles
computational scheme of the complex optical conductivity tensor for
layered systems, without using Kramers--Konig relations, taking into
account, however, both, the inter-- and the intra--band contributions
on the same footing$^{\,7}$.

The computational accuracy is permanently controlled applying the
Gauss--Konrod quadrature and the recently developed cumulative
special--points me\-thod$^{\,8}$. In the present contribution, all the
layer--resolved complex optical conductivity tensors have been
obtained with an accuracy of 0.001 a.u. by taking 35 (2) Matsubara
poles at 300 K in the upper (lower) complex semi--plane and a
life--time broadening of 0.048 Ryd. The Fermi level of -0.038 Ryd
corresponds to that of a paramagnetic fcc--Pt bulk substrate (lattice
parameter of 7.4137 a.u.).

%
\section{Results and discussions}
\label{sect:results}
%
The $2\times2$ matrix formalism has been adopted for polar geometry at
normal incidence, for two reasons, namely (1) Kerr measurements are
mainly recorded under these conditions$^{\,9}$; and (2) in the case of
polar geometry, the solutions of the characteristic equation are
analytically known and hence within the iterative algorithm of
Mansuripur$^{\,2,10}$, the recursion relations are simple form. The
form of $2\times2$ matrices can even be more simplified by neglecting
the difference in the diagonal optical conductivity tensor elements.

Starting from the substrate, once the surface reflectivity matrix has
been iteratively evaluated, the Kerr rotation angle
$\theta_{\mathrm{K}}$ and Kerr ellipticity $\varepsilon_{\mathrm{K}}$,
respectively, are obtained directly by using the following exact
formulas$^{\,11}$
\begin{equation}
\theta_{\mathrm{K}}=-\frac{1}{2}\left(  \Delta_{+}-\Delta_{-}\right)
\ ,
\end{equation}
and
\begin{equation}
\varepsilon_{\mathrm{K}}=-\frac{\left|  r_{+}\right|  -\left|  r_{-}\right|
}{\left|  r_{+}\right|  +\left|  r_{-}\right|  }\ , 
\end{equation}
where $r_{\pm}=\left|  r_{\pm}\right|  e^{i\Delta_{\pm}}$ is the
complex reflectivity of the right-- and  left--handed
circularly polarized light. 

\begin{figure}[hbtp] \centering
\begin{tabular}{c}
\includegraphics[width=0.5\columnwidth,clip]{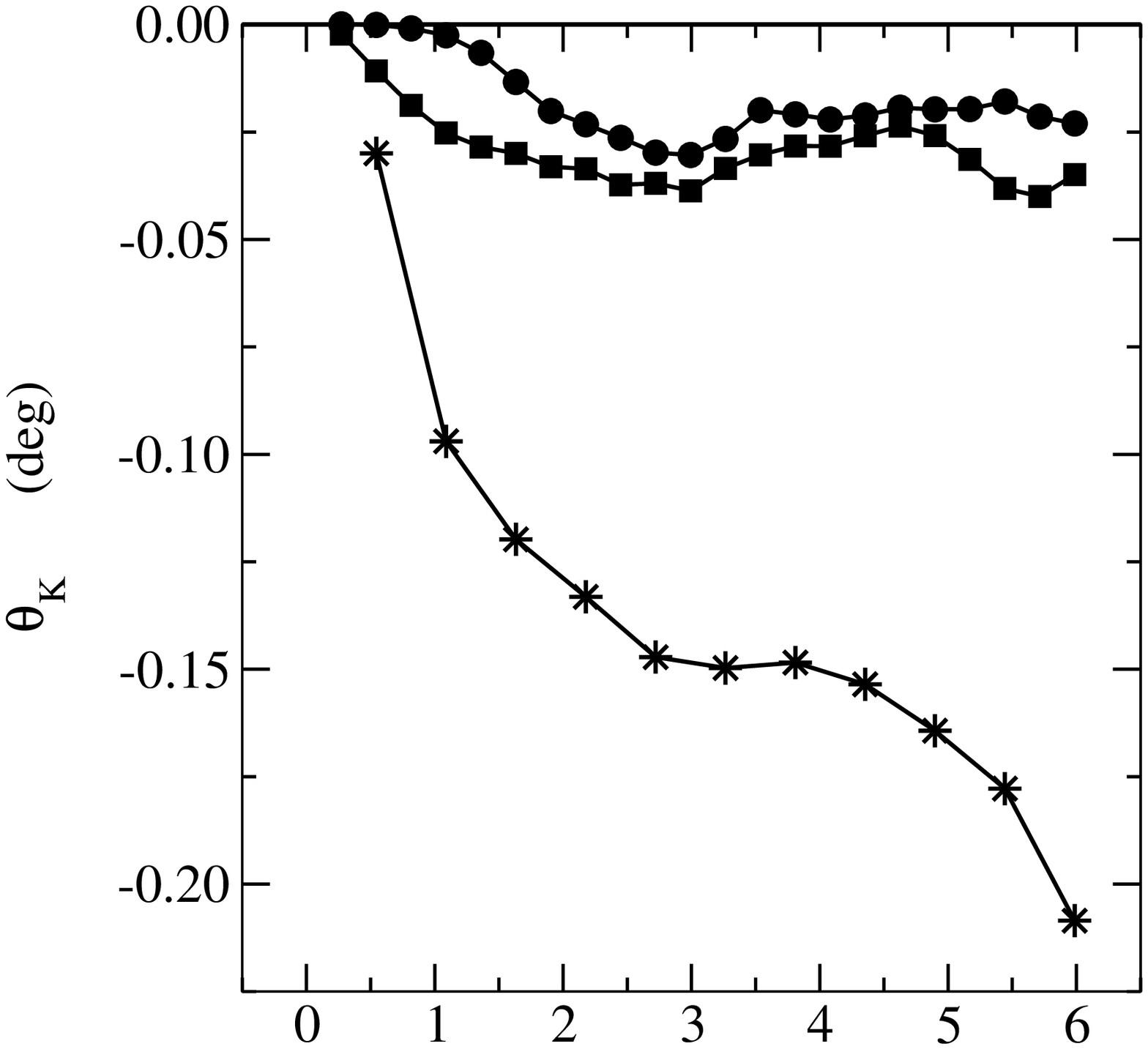} \\
\includegraphics[width=0.5\columnwidth,clip]{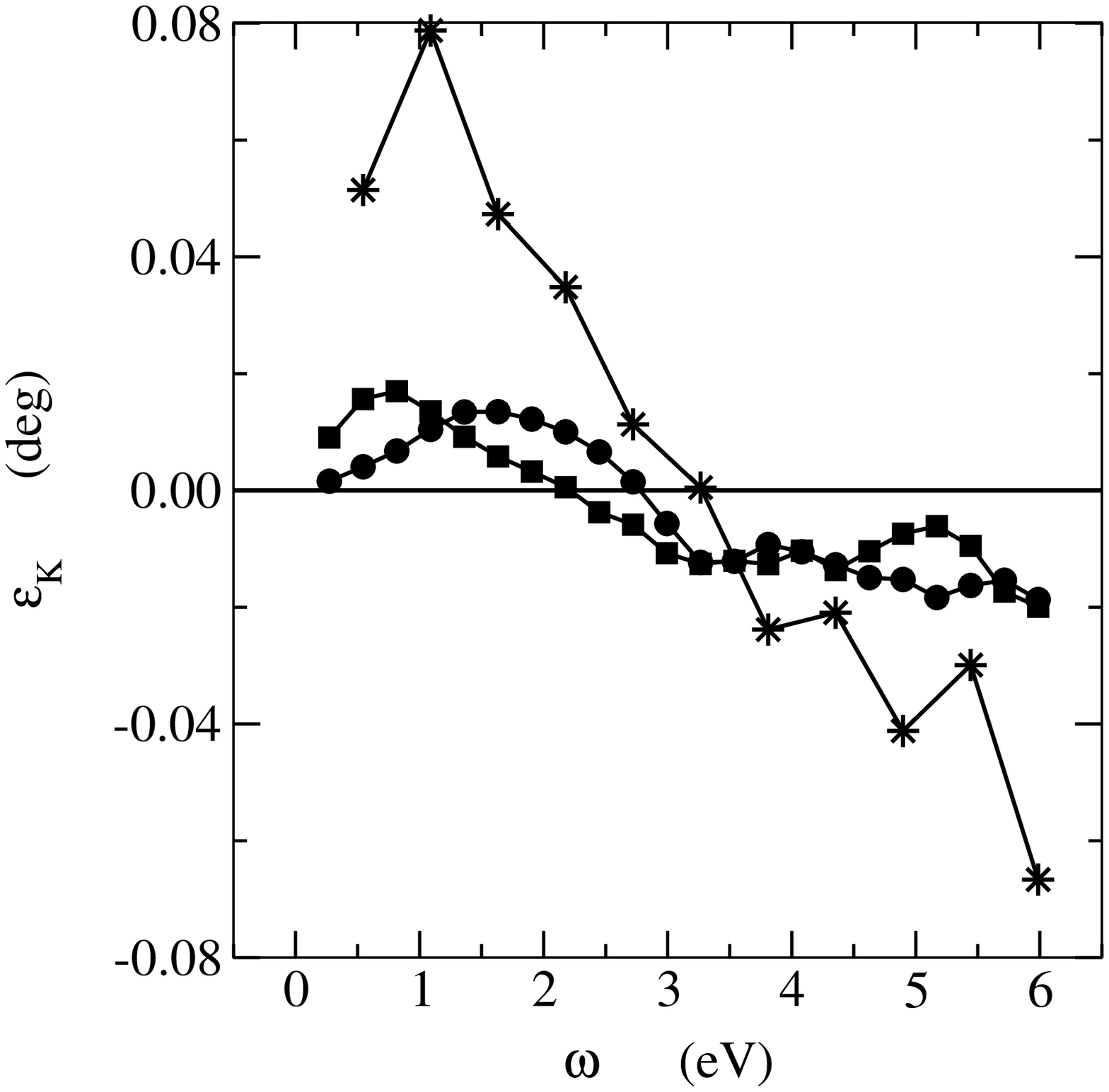} 
\end{tabular}
\caption[]
    {\label{fig:moke-pmsub}
      Kerr rotation angle and ellipticity as a function of the optical
      frequency for fcc(111) $\mathrm{Co}\!\mid\!\mathrm{Pt}_{5}$
      (filled circles), $\mathrm{Pt}_{3} \!\mid\! \mathrm{Co} \!\mid\!
      \mathrm{Pt}_{5}$ (filled squares) and $\mathrm{Pt}_{3} \!\mid\!
      (\mathrm{Co} \!\mid\!  \mathrm{Pt}_{3})_{3} \!\mid\! \mathrm{Co}
      \!\mid\! \mathrm{Pt}_{5}$ (stars), respectively, on top of a
      paramagnetic semi--infinite Pt bulk.  }
\end{figure}

The theoretical Kerr spectra for fcc(111)--$\mathrm{Co} \!\mid\!
\mathrm{Pt}_{5}$ layered system, with three Pt cap layers and in
addition with three $\mathrm{Co} \!\mid\!  \mathrm{Pt}_{3}$ bilayers
in--between the cap layers and $\mathrm{Co}\!\mid\!\mathrm{Pt}_{5}$,
respectively, are shown in Fig.\ \ref{fig:moke-pmsub}. In the
experiments the Pt cap layers on the top of Co layer are deposited to
prevent the oxidation of the surface$^{\,12}$.  We have shown that
strictly speaking the $\mathrm{Co}\!\mid\!\mathrm{Pt}$ multilayer
systems exhibit perpendicular magnetization only in presence of Pt cap
layers. It is therefore not surprising that due to the Pt cap layers
the Kerr effect is enhanced in comparison with uncapped $\mathrm{Co}
\!\mid\!  \mathrm{Pt}_{5}$. The three $\mathrm{Co} \!\mid\!
\mathrm{Pt}_{3}$ bilayers on the top of $\mathrm{Co} \!\mid\!
\mathrm{Pt}_{5}$ increase dramatically the Kerr effect, putting the
theoretical Kerr rotation angle in shape and magnitude in a very good
agreement with the experimental data$^{\,9}$.  However, because the
systems considered here are much smaller than those used in
experiments, a strict quantitative comparison of the Kerr spectra
cannot directly be made.
%
 
%

In conclusion, by combining our technique, which permits to obtain the
layer--resolved optical conductivity tensor on a first principle
basis, with a proper description of the light propagation in
multilayer systems, like the $2\times2$ matrix formalism, a very
realistic description of the magneto--optical Kerr effect can be
given.
%
\section*{Acknowledgements}
%
This work was supported by the Austrian Ministry of Science (Contract
No. 45.451), by the Hungarian National Science Foundation (Contract
No. OTKA T030240 and T029813) and partially by the RTN network
``Computational Magnetoelectronics'' (Contract
No. HPRN--CT--2000-00143).
%

\end{document}